# Machine learning based approach for solving atomic structures of nanomaterials combining pair distribution functions with density functional theory

*Magnus Kløve[a], Sanna Sommer[a], Bo B. Iversen[a]\*, Bjørk Hammer[b]ψ and Wilke Dononelli[c]#*

AUTHOR ADDRESS

[a] Department of Chemistry and iNano, Aarhus University, 8000 Aarhus, Denmark

[b] Center for Interstellar Catalysis, Department of Physics and Astronomy, Aarhus University, Ny Munkegade 120, Aarhus C 8000, Denmark

[c] MAPEX Center for Materials and Processes, Bremen Center for Computational Materials Science and Hybrid Materials Interfaces Group, Bremen University, 28359 Bremen, Germany

AUTHOR INFORMATION

**Corresponding Author**

*Bo B. Iversen: bo@chem.au.dk
ψ Bjørk Hammer: hammer@phys.au.dk
#Wilke Dononelli: wido@uni-bremen.de

## ABSTRACT

Determination of crystal structures of nanocrystalline or amorphous compounds is a great challenge in solid states chemistry and physics. Pair distribution function (PDF) analysis of X-Ray or neutron total scattering data has proven to be a key element in tackling this challenge. However, in most cases a reliable structural motif is needed as starting configuration for structure refinements. Here, we present an algorithm that is able to determine the crystal structure of an unknown compound by means of an on-the-fly trained machine learning model that combines density functional theory (DFT) calculations with comparison of calculated and measured PDFs for global optimization in an artificial landscape. Due to the nature of this landscape, even metastable configurations can be determined.

## TOC GRAPHICS

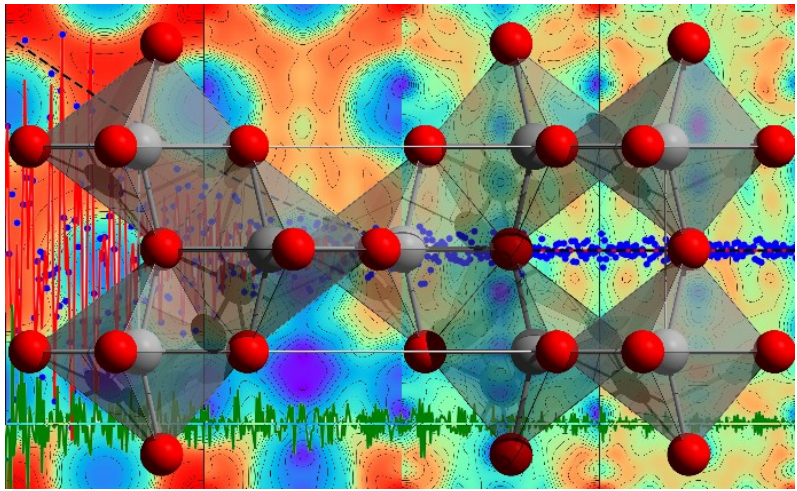

## KEYWORDS

## 1. Introduction

Characterization of functional materials at the atomic level plays a crucial role in the advancement of modern society, as the properties of functional materials are intimately related to their atomic structure[1–3].

In this regard, X-ray and neutron scattering techniques are indispensable tools, since they are direct probes of the atomic arrangement in solids, and the atomic structure of unknown compounds can be routinely elucidated from 3D single-crystal diffraction (SCD) data[4]. For a powdered sample, however, the 3D structural information condenses onto a 1D diffractogram causing significant peak overlap and loss of directional information. Still, phase retrieval procedures such as the Patterson method and Direct Methods have been applied for solving atomic structures from powder X-ray diffraction (PXRD) experiments analogous to structure solution from SCD data[5–8]. Global optimization techniques such as genetic algorithms[9,10], Monte-Carlo optimization[11–14] and simulated annealing[11,15] attempt direct-space atomic structure solution from agreement with experimental PXRD data utilizing raw computational power. Hybrid methods such as the works of Wolverton and coworkers on FPASS build upon the same idea of global optimization using a genetic algorithm, however, with the use of density functional theory (DFT) energies to identify low energy structural motifs during the search[16,17]. In all cases, however, structural solution from PXRD typically requires a homogenous and well-crystalline sample together with high quality data.

Real, functional materials are not necessarily well-crystalline, but are likely to be nanostructured, and so the traditional crystallographic approaches to structure solution fall short. For these materials, the concept of an infinitely repeating unit cell breaks down. This results in peak broadening and thus even more severe peak overlap in the 1D diffractograms, yet again reducing the amount of unique information available[18,19].

The pair distribution function (PDF) analysis of X-ray or neutron total scattering data has proven a valuable tool for studying nanomaterials[20–23]. Still, most PDF analyses are guided by *a priori* structural information and are carried out as real-space counterparts to Rietveld refinements. Recent efforts by the Billinge group on the cloud platform, *PDFitc.org*, have provided the PDF community with real-space matching software for identifying unknown phases in PDF patterns through a structural database search[24,25]. While such software can aid real-space analyses if the structure has been previously determined, determining the atomic structures of truly unknown nanocrystalline phases from PDF data remains yet a pending task for the community in the absence of efficient and versatile procedures.

Large-box global optimization like Reverse Monte-Carlo (RMC) modelling of PDFs has long been applied to study the atomic structure of amorphous materials, where no unique structural solution exists, but where the statistical average of the ensemble can provide structural insight[26,27]. However, the RMC algorithm is inherently inefficient due to its unguided, random-guessing approach, often leading to search trapping in local minima.

The "Liga" method originally showcased to solve the structure of single-element, non-periodic clusters like $C_{60}$ from experimental PDFs[28] has also been extended to solve atomic structures of multiple-element crystalline samples from experimental PDFs[29]. While successful in most cases, the algorithm has only been tested on well-crystalline powder samples, where traditional reciprocal space analysis could be more intuitively applied.

In this letter, we present an efficient machine-learning enhanced global optimization algorithm for solving atomic structures from experimental PDFs of materials on several length scales including nanomaterials, obtained with only short temporal resolution. By combining the global optimization method GOFEE by Bisbo and Hammer[30,31] with the DiffPy-CMI software[32] for PDF modeling, the algorithm performs a global optimization in an on-the-fly, machine learning trained surrogate R-factor landscape, thus solving the atomic structure from the PDF. Combining DFT calculations with PDF modelling the machine learning model can be trained in a combined artificial R-factor/DFT landscape to automatically determine the structure of metastable phases correctly.

## 2. Combining R-factor landscape and DFT Energy for global optimization.

In the original work of Bisbo and Hammer, the GOFEE algorithm was used to optimize structures to their global minimum in the potential energy surface (PES) calculated using density functional theory (DFT). For a detailed description of the GOFEE method, readers are referred to the original work by Bisbo and Hammer[30,31]. In this work, we want to present an alternative approach to a pure DFT search, where the global optimization is additionally guided based on agreement between a calculated and an experimentally measured PDF through the R-factor in Eq. (1). A machine learning potential is trained on-the-fly on this comparison function and performs local optimizations of new candidate structures in this artificial (R-factor) landscape in combination with the DFT PES (see Eq. (2)), eventually optimizing the structures to their global minimum after several iterations of the algorithm.

$$R_w = \sqrt{\frac{\sum_i w_i(G_{i,obs} - G_{i,calc})^2}{\sum_i w_i G_{i,obs}^2}} \qquad (1)$$

In Eq. (1), the R-factor is defined as the deviation between the calculated PDF, $G_{i,calc}$, and the observed PDF, $G_{i,obs}$, at each data point $i$ with the weights, $w_i$, defined as $1/\sigma_i^2$., with $\sigma_i$ being the uncertainty of $G_{i,obs}$. In the underlying work, these R-factors are determined using the DiffPy CMI software[32].

The fitness F of new candidate structures within the GOFEE framework is calculated by combining the R-factor with prefactor $\alpha$, with the total energy of the system $E_{DFT}$ (in units of eV) calculated at DFT level of theory with prefactor $(1 - \alpha)$ as shown in Eq. (2).

$$F = \alpha \cdot R_w + (1 - \alpha) \cdot E_{DFT} \qquad (2)$$

Eq. (2) allows different mixing factors for $R_w$ and the total DFT energy to be used as the fitness function in the global optimization run. The two extrema correspond to setting either $\alpha = 0$ or $\alpha = 1$. In the first case, the global optimization is only based on the DFT energy, in the latter the search is only guided based on agreement with the PDF through $R_w$.

## 3. Results and Discussion

### 3.1 Finding the crystal structure of 6 nm Anatase nanoparticles

To demonstrate the ability of the algorithm to find the crystal structure of an unknown compound and to analyze the influence of the prefactor $\alpha$, we globally optimize the atomic structure of 6 nm Anatase ($TiO_2$) nanoparticles from a PDF obtained with only 3 s of X-ray exposure. Each run is initialized by creating four different random structures by placing 4 Ti and 8 O atoms at random non-overlapping positions in the simulation cell. Each of the four initial candidates is then locally optimized for 10 BFGS steps. After the initialization and local optimization, 1500 GOFEE cycles are used to guarantee convergence of the algorithm. Details on the algorithm can be found in the methods section and for details on GOFEE in general in the original work by Bisbo and Hammer[30,31].

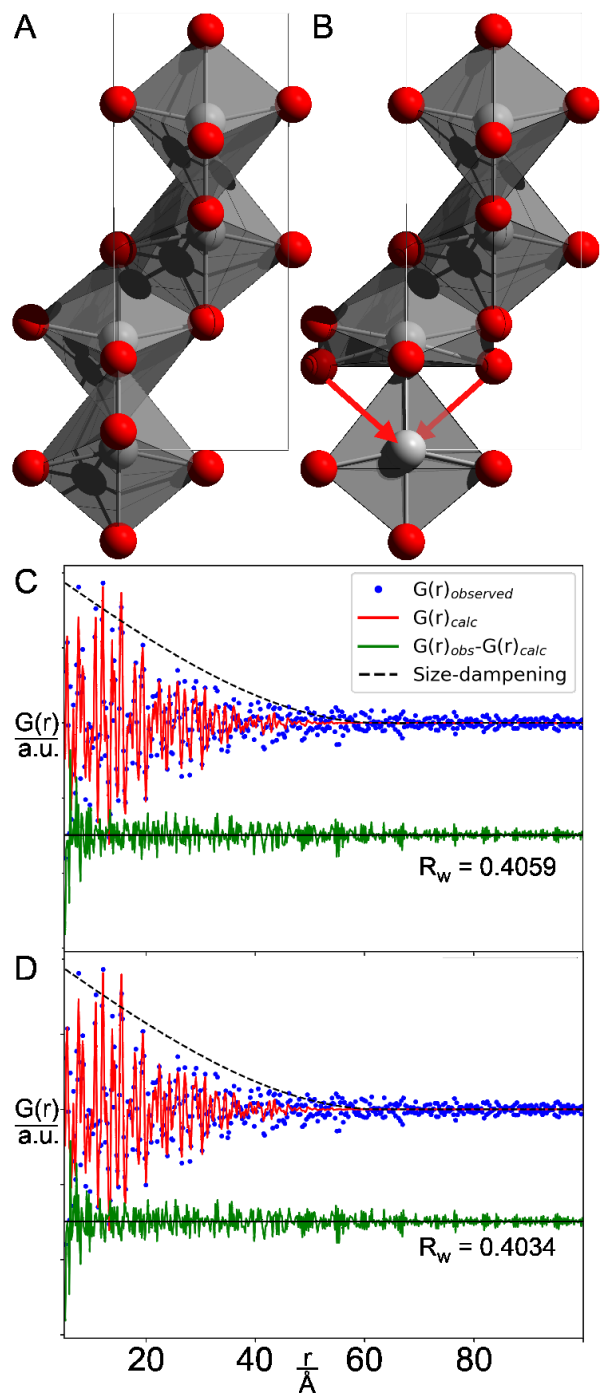

**Figure 1:** Result of global optimization based on R-factor compared to reference structure as found by DFT. A: reference structure; B: resulting structure of global optimization only based on R-factor ($\alpha = 1$). Both structures have been postprocessed by local optimization based on the R-factor. The red arrows in B indicate the correct position of the displaced O atom with respect to the non-defected crystalline structure as shown in A; C: PDF pattern of of perfectly crystalline structure in A; D: PDF pattern of structure B. Grey atoms are titanium and red atoms are oxygen.

Using the R-factor alone ($\alpha = 1$), the global optimization yields a best-fit structure where one oxygen atom is displaced with respect to the perfectly ordered crystalline structure (cf. Figure 1 A and B). As shown in Figure 1 C and D, the difference in the corresponding $R_w$ values is only very small (0.025). Nevertheless, such a small difference is enough to force the algorithm to focus on structures that show such a structural defect.

To analyse this effect in a statistical fashion, ten different instances of the same global optimization were started for four different values of $\alpha$, respectively. The results are shown in Figure 2 B. GOFEE was able to determine the correct atomic arrangement, i.e. crystal structure as shown in Fig. 2 A, of non-defected $TiO_2$ from the PDF ($\alpha = 1$) with a success rate of 60% (c.f. Figure 2 B red line). The criterion for success is defined as finding a structure in the pool of candidates that measured by comparison of the sum of all bond lengths agrees with a corresponding non-defected crystal structure as shown in Fig. 2A. We found a similar result for $CeO_2$ as shown in the Supporting Information, although due to higher symmetry of the $CeO_2$ crystal structure, the success rate was found to be 80%. In contrast to global optimization of the Anatase structure based on the R-factor ($\alpha = 1$), a global optimization based entirely on the DFT energy ($\alpha = 0$) gives a 100% success rate already after a few (54) iterations, however, without considering the agreement with experimental data (the $R_w$ value) during the search. Using values of $\alpha = 0.5$ and $\alpha = 0.8$ leads to successrates inbetween the ones of the two extreme cases. Note: a value of $\alpha = 0.5$ does not necessary lead to an equal wighting of the DFT energy and the $R_w$ values. $R_w$ ranges from 0 to 1, whereas the total energy of a system $E_{DFT}$ might differ by orders of magnitude for depending on chemical composition and atomic arrangement.

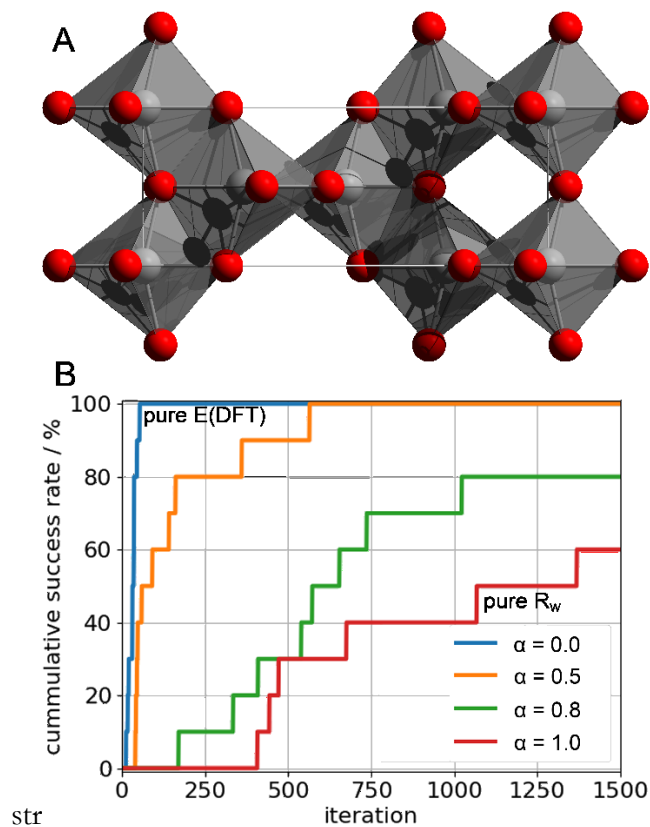

**Figure 2:** Result of global optimization based on R-factor, DFT energy and combination of both. A: Non-defected reference structure defining a successful run; B: cumulative success rate vs. number of GOFEE iterations for finding a structure similar to structure A based on different values of α. Grey atoms are titanium and red atoms are oxygen.

Now, the question arises why the global optimization based on the DFT energy is much more efficient than the same search based on the R-factor. The answer to this question is illustrated in Figure 3, where the landscapes for placing the shaded Ti atom at every position in the (100)-plane is shown for four combinations of DFT/PDF weighting. The pure DFT PES shows five maxima (blue) and just three distinct favourable minima (red). In comparison, the corresponding pure R-factor landscape shows multiple different minima. Exploration of all these minima takes much more effort than exploring the few local minima present in the DFT PES. Combining the DFT energy with the R-factor yields intermediate landscapes with fewer or at least more pronounced local minima (Figure 3) and consequently leads to success rates in between the ones of a pure DFT or R-factor based search (Figure 2B). As a side effect, even for weighting the R-factor stronger than the DFT energy (Figure 2B, green line), the somewhat unphysical features of the pure R-factor based search can be discarded and a structure similar to the perfect, non-defected anatase crystal structure results.

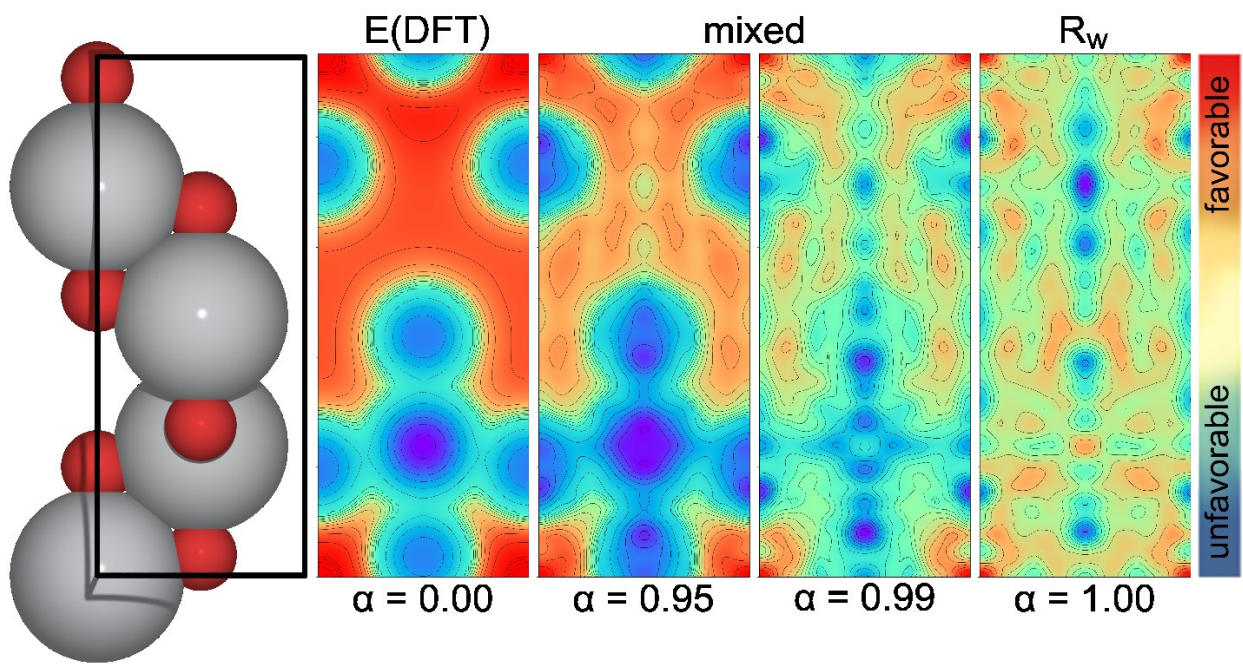

**Figure 3:** Normalized E(DFT) potential energy landscape, R-factor landscape and combined E(DFT) and R-factor landscapes for Anatase. The transperent Ti atom is placed at all positions in the unit cell for its given correct height. Corresponding DFT Energies and $R_w$ values are calculated for each point. α values differ from values used in Fig. 1 for illustrational puposes. Grey atoms are titanium and red atoms are oxygen.

## 3.2 Studying the defected crystal structure of Gahnite $ZnAl_2O_4$

The R-factor landscape of solid matter can become quite complex, as demonstrated for Anatase resulting in only 60% success rate in a pure R-factor based search for this rather simple crystal structure. In the case of Gahnite ($ZnAl_2O_4$), which crystallizes in the more complex spinel structure that features a large unit cell with both octahedrally and tetrahedrally coordinated metal ions, the R-factor landscape seems to be too complex for the GOFEE algorithm to determine a physical meaningful average structure only based on the R-factor even after many iterations of the global optimization algorithm (see Figure S2). In the R-factor optimized structure, some local structural motifs of the metal ions that resemble the octahedral and tetrahedral coordination found in the spinel crystal structure are present. Yet, the overall structure is far from being a realistic representation of the crystal structure. Using a pure DFT PES ($\alpha = 0$), the GOFEE algorithm yields the perfect, non-defected crystal structure as known from databases. Depending e.g. on the functional used for energy evaluations (e.g. a pure GGA like PBE or a hybrid functional like PBE0), some distances and angles between different atoms might differ slightly, however, the overall motif still matches the spinel structure.

Regardless, both the R-factor and DFT optimized structures provide fits to the experimental PDFs of similar quality, and this example therefore clearly shows the limitations of using GOFEE or any other global optimization method weighted only by the R-factor, whenever the structure is more complex than a simple, highly symmetric binary compound.

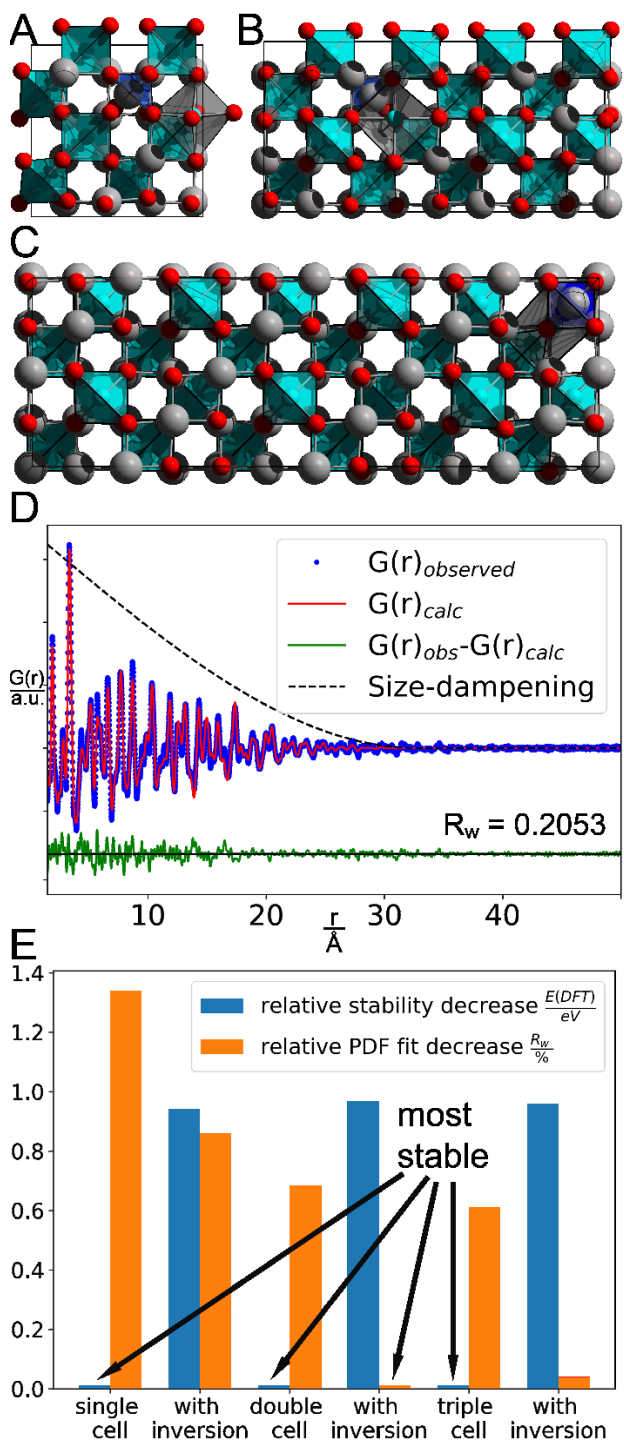

**Figure 4:** Modeled PDF and structure of Gahnite global optimized based on the R-factor and DFT energy;.Inversion in a single unit cell; B: Inversion in a double unit cell; C: Inversion in a triple unit cell; All three models were predicted by GOFEE in the combined DFT PES and R-factor landscape; D: computed PDF of structure A with corresponding $R_w$ value ; E: stability/likelihood comparison of Gahnite unit cells with and without inversion. Grey atoms are aluminum, blue atoms are zinc and red atoms are oxygen.

If on the other hand, the R-factor and DFT energy are combined in the global search, a different structure was found showing an inversion of one Al atom with one Zn atom (Figure 4A). This is a well-known point defect in spinel structures[33,34] and therefore not too surprising. Nevertheless, the locally optimized (based on the DFT energy) non-defected structure is 0.93 eV more stable than the locally optimized defected structure. On the other hand, the calculated PDF of the defected structure (locally optimized using numerical R-factor forces) shows 0.63 % better agreement to the meassured PDF than the calculated PDF of the locally optimized non-defected structure (c.f. Fig. 4 E). Therefore, although the non-defected structure might represent the thermodynamically most stable configuration best, the defected structure found with the combination of DFT energy evaluation and R-factor calculation, leads to the best average representation of the studied Gahnite nanoparticles with respect to the meassured PDF pattern.

In the above, we worked with a 1x1x1 cell and hence a high concentration of defects whenever these are present. In an attempt to study the occurrence of defects at lower concentrations, a global optimization of a 2x1x1 Gahnite unit cell was run and similar results were obtained. The search based on the DFT energy leads to a non-defected crystal structure, whereas the search in the combined DFT and R-factor landscape leads to a structure with an inversion of Zn and Al. Comparable to the single unit cell, the non-defected structure is 0.95 eV more stable. On the other hand, again, locally optimizing the resulting structures in the R-factor landscape leads to 0.86 % better fit of the PDF of the defected structure. This is slightly more pronounced that in the case of a single unit cell. Increasing the cell size to a 3x1x1 super cell, the non-defected structure is 0.96 eV more stable than the one showing one inversion comparable to a single unit cell and a 2x1x1 super cell. On the other hand the defected super cell yields now only a 0.72% better fit than the non-defected structure. Therefore, one can estimate that the concentration of point defects is likely lower than one Al/Zn inversion per unit cell, but bigger than 0.33 inversions per unit cell. Note, however, that in the case of the 3x1x1 super cell the dimension of the periodically repeated cell almost reaches the size of the nanoparticle itself. Therefore, some additional effects might influence the decrease in the stability difference. Finally it should be noted that we have considered a perfect stoichiometry in the search, whereas non-stoichiometric amounts of the three elements were used in the refinement in the original study by Sommer et al.[34] from which the data originate. Using non-stoichiometric amounts of any of the three elements in our GOFEE search could potentially result in other structural motifs showing up such as the different point defects found in the their work.

## 4. Conclusion and outlook

We have introduced a global optimization method, which utilizes methods from the field of machine learning to find the crystal structure of unknown compounds only based on the comparison of a modeled to a measured PDF (based on the agreement factor, $R_w$). This method alone works well for simple crystal structures, but for more complex materials, one needs to combine the R-factor with an energy evaluation of the system thus introducing chemical intuition in the algorithm, e.g. by means of DFT in order to sort out physical non-meaningful structures. In addition, the combined DFT+R-factor corrected landscape may lead to metastable configurations as global minima, whereas a search based on DFT would always try to predict the thermodynamically most stable configuration as the global minimum. Using this method, crystal structures of unknown compounds can be solved fully automatically based on a combination of PDF measurements and DFT calculations within only a few hours to days of computational time. Although only tested with combination of the DFT energy with the R-factor, the method is not limited to combine these two values and create an artificial landscape for the global optimization. As alternatives classical potentials or force fields could potentially be used in order to force the algorithms to generate physically meaningful structures. We additionally assume that instead of using the $R_w$ factor from a PDF agreement evaluation, other experimental comparisons could be used. One example could be to utilize the R-factor from a comparison of a measured and a calculated PXRD in combination with DFT, or even in a combined DFT+R(PDF)+R(PXRD) landscape evaluation.

5. Methods

**5.1 Experimental methods.** The TS data on $TiO_2$ and $ZnAl_2O_4$ was collected at the P02.1 beamline at PETRA-III (DESY, Hamburg, Germany) with a photon energy of ~60 keV ($\lambda$ = 0.2072 Å)[35] on a 2D PerkinElmer XRD 1621 area detector for data acquisition. The TS data on $CeO_2$ shown in Supporting Information was collected at the P07 beamline at PETRA-III (DESY, Hamburg, Germany) with a photon energy of ~100 keV ($\lambda$ = 0.124 Å)[36] on a DECTRIS PILATUS3 X 2M CdTe detector. The 2D detector images were azihmutally integrated with the Dioptas[37] software and calibrated using either a $LaB_6$ 660b NIST line standard for $TiO_2$ and $ZnAl_2O_4$ or the $CeO_2$ data itself. The data was appropriately background subtracted and subsequently Fourier transformed using PDFgetX3[38] in the xPDFsuite[39] software, thus giving the PDF.

The TS data of Anatase ($TiO_2$) nanoparticles was obtained from an *in situ* synthesis by Søndergaard-Pedersen et al.[40] An industrial grade $TiOSO_4$ precusor was diluted with sulfuric acid to a titanium concentration of 1.29(1) M. The solution was injected into a fused silica capillary mounted in a revised version of the *in situ* setup described in detail elsewhere[41], pressurized to 250 bar and heated to 200 °C. Data collection was carried out with a temporal resolution of 3 s. The data was background subtracted with a measurement of diluted sulfuric acid and Fourier transformed with a $Q_{max}$ = 16.8 Å$^{-1}$. Instrumental dampening was accounted for by separately obtaining and refining TS data of a LaB$_6$ 660b NIST line standard in the PDFgui[42] software.

The TS data of Gahnite (ZnAl$_2$O$_4$) nanoparticles originates from a study by Sanna Sommer et al.[34], and readers are referred to their paper on the preparation of the nanoparticles. The data was background subtracted with a measurement of an empty capillary and Fourier transformed with a $Q_{max}$ = 23.5 Å$^{-1}$. Instrumental dampening was accounted for by separately obtaining and refining TS data of a LaB$_6$ 660b NIST line standard in the PDFgui[42] software.

The TS data of CeO$_2$ was background subtracted with a measurement of an empty capillary and Fourier transformed with a $Q_{max}$ = 25.0 Å$^{-1}$. Instrumental dampening was account for by performing an initial refinement of the data in the TOPAS[43] v6 software.

**5.2 Computational methods.**

*Local structure optimization*

Local geometry optimizations can be performed straightforwardly for DFT calculations using analytical forces as implemented in GPAW[44]. The exchange-correlation interaction was treated by the generalized gradient approximation using the PBE functional.

In addition, local geometry optimizations can be performed in the articifial R-factor landscape. Instead of using an analytical expression, forces are calculated numericaly by displacing the atom positions ± 0.001 Å into each Cartesian direction and calculating the corresponding R-factor (from Eq. (1)). Then, the BFGS algorithm as implemented in ASE[45] is used to optimize the structure into its local minimum.

In this work, the global optimization method GOFEE is used. In contrast to the original work by Bisbo and Hammer[30,31], the machine learning potential is not always exclusively trained on DFT energies, but additionally on the R-Factor (as defined in Eq. (1)) or a combination of the R-factor and DFT energy according to Eq. (2).

The workflow of the code can be divided into five steps. First (I), a user defined number of initial random atomic structures is created, e.g. 10 structures. All structures share a common set of lattice parameters and stoichiometry. Each of these structures are geometrically locally relaxed either using DFT forces or numerical R-factor forces. Based on this data, the initial machine learning potential can be trained. In the second step (II), $m$ new structures are created via random guess or mutation of the previous population ($m$ = number of CPU cores). In the current version of the code, the mutations are either permutation of two non-identical atoms or rattling some of the atoms to new positions. In the third step (III), all new structures' atomic positions are relaxed in the model machine learning potential. Next (IV), the fitness of each of these new structures is evaluated in the machine learning potential. Then a structure, where the machine learning potential either predicts a good fitness (exploit) or has a high uncertainty (explore), is choosen. For this structure, the correct fitness is evaluated. The fitness can either be the DFT potential energy, the R-factor as introduced in Eq. (1) or a combination of both as introduced in Eq. (2). In the last step (V), the chosen new structure and its corresponding fitness is added to training data and the model potential is retrained. The algorithm starts again from step (II) until $n$ iterations has been reached (e.g. $n$ = 2000), resulting in hundreds of different local minimum structures. The resulting structures including the global minimum structure were only relaxed in the ML potential during the search. Therefore, the resulting global minum structure has to be post-processed and has to be locally optimized as described above. If the energetic stability should be compared, the final structure is relaxed using analytical DFT forces. If the fit to the experiment is important, then numerical R-factor forces are used for the final local relaxation.

**Supporting Information**

The Supporting Information is available free of charge at:

Experimental section, algorithm description, additional figures (PDF)

**Notes**

The authors declare no competing financial interests.

**Acknowledgement**

Funding

We acknowledge financial support through the APF project 'Materials on Demand' within the 'Humans on Mars' Initiative funded by the Federal State of Bremen and the University of Bremen and support from the the Villum Fonden. In addition, this work was supported by the Danish National Research Foundation through the Center of Excellence "InterCat" (Grant Agreement No. DNRF150)